\documentclass[12pt,reqno,a4paper]{article}
\usepackage{amsmath,amsfonts,amssymb,graphicx,epsfig,pict2e,rotating}
\usepackage[english]{babel}
\usepackage[noxcolor]{pstricks}
\usepackage{hyperref}
\numberwithin{equation}{section}

\newcommand{\be}{\begin{equation}}
\newcommand{\ee}{\end{equation}}
\newcommand{\bea}{\begin{eqnarray}}
\newcommand{\eea}{\end{eqnarray}}
\newcommand{\D}{{\mathbf D}}

\newcommand{\uno}{{\mathbf 1}}
\newcommand{\T}{{\mathbf T}}
\newcommand{\e}{{\mathbf e}}

\begin{document}
\setcounter{page}{1}

\vspace{8mm}
\begin{center}
{\Large {\bf Lattice Integrals of Motion of the Ising Model on the Cylinder  }}

\vspace{10mm}
 { Alessandro Nigro\\
 Dipartimento di Fisica and INFN- Sezione di Milano\\
 Universit\`a degli Studi di Milano I\\
 Via Celoria 16, I-20133 Milano, Italy\\
 Alessandro.Nigro@mi.infn.it}\\
 [.4cm]

  \end{center}

\vspace{8mm}
\centerline{{\bf{Abstract}}}
\vskip.4cm
\noindent
We consider the 2D  critical Ising model with spatially periodic boundary conditions. It is shown that for a suitable reparametrization of the known Boltzmann weights the transfer matrix becomes a polynomial in the variable $\csc(4u)$, being $u$ the spectral parameter. The coefficients of this polynomial are decomposed on the periodic Temperley-Lieb Algebra by introducing a lattice version of the Local Integrals of Motion. 

\renewcommand{\thefootnote}{\arabic{footnote}}
\setcounter{footnote}{0}
\section{Introduction}
In this paper we push our investigation of the critical Ising model a bit farther. Our objective is to extend the results of the previous work on Critical Dense Polymers \cite{nigropol} to the Ising work, in order to put the previous work on a more solid ground and to show that once more the same approach is not only possible, but can in this case be carried out completely.\\
The approach goes all the way back to \cite{blz}, where the integrable structure of conformal field theory (CFT) was analyzed from the point of view of the existence of infinitely many higher spin conserved quantities in mutual involution. Inspired by this work we seek and succeed to obtain the complete tower of commuting conserved quantities.\\
The involutive charges of CFT are defined as the maximal Abelian subalgebra of the enveloping algebra $UVir$ of the Virasoro algebra, the lattice analogue of this structure is obtained by replacing the Virasoro algebra with the Temperley-Lieb algebra, as it is well known \cite{saleur}.\\
There are of course different types of Temperley-Lieb (TL) algebras, in particular it makes a lot of difference to work with open boundaries or periodic boundaries. In \cite{nigropol} we worked with an open boundary TL algebra, and obtained closed form expressions for many of the involutive charges, an ansatz closed form for the generic expressions was also provided in which some of the coefficients are still unfortunately not known. The heart of the difficulty, as pointed out long ago also in \cite{saleur} lies in the presence of boundary terms which manifest themselves through the appearence of boundary tangles. It is therefore natural to expect, and this is indeed the case, that such difficulties can be overcome by considering a model based on a periodic TL algebra. In this case there are no boundary tangles and the expressions for the involutive quantities simplify enough that a closed form for them can be found. This is surely the case for Critical Dense Polymers on the Cylinder \cite{cylpol} as well, since it shares an extremely large number of analogies with the Ising model among which are almost identical expressions for inversion identities, eigenvalues of the transfer matrix and expression for the continuum involutive charges.\\

\section{Generalities}
It is well known that the Ising model can be characterized in terms of Boltzmann weights, which can be derived from an elementary lattice hamiltonian \cite{ret1}. Naturally such weights can be suitably rescaled by multiplying for trigonometric funtions, in particular we shall employ here the following parametrization:
\begin{equation}\begin{split}W_R\left(\left.\begin{matrix}\rho&\\[-2mm]&\tau\\[-2mm]\sigma&\end{matrix}
\right| \,u \right)= \Big(\tan\Big(\frac{\pi}{4}-u\Big)\delta_{\sigma,\rho}+\delta_{\sigma,-\rho}\Big)\delta_{\sigma,\tau}+\Big(\cot\Big(\frac{\pi}{4}-u\Big)\delta_{\sigma,\rho}+\delta_{\sigma,-\rho}\Big)\delta_{\sigma,-\tau} \end{split} \end{equation}
\begin{equation}\begin{split}W_L\left(\left.\begin{matrix}&\rho\\[-2mm]\tau&\\[-2mm]&\sigma\end{matrix}
\right| \,u \right)= \Big(\cot(u)\delta_{\sigma,\rho}+\delta_{\sigma,-\rho}\Big)\delta_{\sigma,\tau}+\Big(\tan(u)\delta_{\sigma,\rho}+\delta_{\sigma,-\rho}\Big)\delta_{\sigma,-\tau}  \end{split} \end{equation}
satisfying
\be W_R\left(\left.\begin{matrix}\rho&\\[-2mm]&\tau\\[-2mm]\sigma&\end{matrix}
\right| \,\frac{\pi}{4}-u \right)=W_L\left(\left.\begin{matrix}&\rho\\[-2mm]-\tau&\\[-2mm]&\sigma\end{matrix}
\right| \,u \right)   \ee
A transfer matrix can then be defined in different ways and for different boundary conditions, in particular if we choose to work on the cylinder we can build the transfer matrix in a boundary condition dependent way evolving into periodic imaginary time, or on the contrary we can choose to build a boundary indipendent transfer matrix with spatially periodic boundaries which evolves in non compact time. We choose here to explore the second possibility, the first one having been already exhaustively discussed in \cite{ret1} and also elsewhere.\\
Therefore we introduce the periodic transfer matrix:
\be T_{\mathbf{ \sigma},\mathbf{ \rho}}(u)=\sum_{{\bf \tau}}\prod_{n=1}^L W_L\left(\left.\begin{matrix}&\rho_n\\[-2mm]\tau_n &\\[-2mm]&\sigma_n\end{matrix}
\right| \,u \right)W_R\left(\left.\begin{matrix}\rho_n &\\[-2mm]&\tau_{n+1}\\[-2mm]\sigma_n&\end{matrix}
\right| \,u \right)   \ee
with 
\be \tau_{n+L}=\tau_n   \ee
and similar conditions for $\sigma, \rho$. The transfer matrix has thus size $2^L\times 2^L$.\\
The partition function is then defined as:
\be  Z=\textrm{Tr}\T^M   \ee
on general grounds the partition function is expected to behave asymptotically as a linear combination of CFT characters, whose modular parameter is related to the exponential of some trigonometric function of the spectral parameter. We will describe this behaviour of the partition function in a later section.\\
It is also well known that the Ising model and in generally Spin-chains, Potts models, RSOS models, Vertex models \cite{saleur} are directly related to representations of the Temperley-Lieb (TL) algebra. We will extensively investigate this connection for this special case, thus for our needs it is necessary to introduce the periodic TL algebra, whose generators are defined by the following matrix elements: 
\be (\e_{2i})_{\mathbf{\sigma},\mathbf{\rho}}=\sqrt{2}\delta_{-\sigma_i,\sigma_{i+1}}\prod_{k=1}^L \delta_{\sigma_k,\rho_k}  \ee
\be (\e_{2i-1})_{\mathbf{\sigma},\mathbf{\rho}}=\frac{1}{\sqrt{2}}\prod_{k\neq i}\delta_{\sigma_k,\rho_k}  \ee
which satisfy the relations:
\be \e_i^2=\sqrt{2}\e_i  \ee
\be  \e_i\e_{i\pm 1}\e_i=\e_i      \ee
\be \e_i\e_j=\e_j\e_i \ , \ |i-j|\geq 2  \ee

\section{Inversion Identity and Eigenvalues}
In this setion we introduce a quadratic functional equation satisfied by the transfer matrix $\T$, the techniques here employed naturally follow the work \cite{ret1}, needless to say. We call it Inversion identity, even though strictly speaking its right hand side is not proportional to the identity, nonetheless this is the historical name introduced by Baxter in his book \cite{baxter} in a very similar case and therefore we shall use it.\\
We also drop for the rest of the paper the use of trigonometric functions of the spectral parameter $u$ in favour of the extremely more convenient variable:
\be x(u)=\sin(4u)  \ee
notice that
\be  x\big(u+\frac{\pi}{4}\big)=-x(u)  \ee 
One has then that the transfer matrix satisfies the following functional equation:
\be  \T(x)\T(-x)=(-1)^L 2^{2L+1}\big(T_{2L}(1/x)\uno+\mathbf{R}\big)  \ee
where $T_n(z)$ are the Chebyshev polynomials of the 1st kind, defined by:
\be T_n(\cos\theta)=\cos(n\theta)  \ee
and ${\bf R}$ is an involution
\be {\bf R}^2=1   \ee
which also commutes with the transfer matrix $\T$:
\be [\T(x),\mathbf{R}]=0 \ee
it is thus possible to diagonalize the involution and the transfer matrix simultaneously, and, calling  $r=\pm 1$ the eigenvalue of the involution, and $\Lambda(x)$ those of the transfer matrix, one has:
\be \Lambda(x)\Lambda(-x)=(-1)^L 2^{2L+1}\big(T_{2L}(1/x)+r\big)  \ee
which can be factorized as:
\be   \begin{split} \Lambda(x)\Lambda(-x)=&(-1)^L 2^{4L}\prod_{k=1}^{2L}\Bigg(\frac{1}{x}+\sin\Big(\frac{2k-\delta_{r,1-2\textrm{Mod}(L,2)}}{2L}\pi\Big)\Bigg)\\
=&(-1)^L 2^{4L}\prod_{k=1}^{L}\Bigg(\frac{1}{x^2}-\sin^2\Big(\frac{2k-\delta_{r,1-2\textrm{Mod}(L,2)}}{2L}\pi\Big)\Bigg) \end{split}      \ee
which sharing out the zeroes leads to the following factorized form of the eigenvalues:
\be  \Lambda(x,\mathbf{\mu},r,L)=\frac{2^{2L}}{x^L}\prod_{k=1}^L\Bigg(1-\mu_k x \sin\Big(\frac{2k-\delta_{r,1-2\textrm{Mod}(L,2)}}{2L}\pi\Big)\Bigg)  \ee
where:
\be \mathbf{\mu}=(\mu_1,\ldots,\mu_L)   \ee
\be \mu_k^2=1  \ee
given ${\bf \mu}$ we can also introduce the associated Fermionic Partition $\mathcal{P}=(k_1,\ldots,k_p)$ ($p\leq L$), defined by:
\be k\in\mathcal{P} ,\ \textrm{if}\ \mu_k=1  \ee
for example $\mathbf{\mu}=(1,-1,-1,-1,1,1,-1,1)$ is associated to the partition $\mathcal{P}=(1,5,6,8)$.\\
Now, by calling $\Lambda_0^r$ the eigenvalue corresponding to ${\bf \mu}=(1,1,1,1,1,\ldots)$,and introducing the auxiliary functions:
\be p_k^r(x)=\frac{1+x\sin\Big((2k-\delta_{r,1-2\textrm{Mod}(L,2)})\frac{\pi}{2L}\Big)}{1-x\sin\Big((2k-\delta_{r,1-2\textrm{Mod}(L,2)})\frac{\pi}{2L}\Big)}  \ee
one has that the generic eigenvalue is decomposed as:
\be \Lambda^r(x)=\Lambda^r_0(x)\prod_{k\in\mathcal{P}}p_k^r(x)   \ee
we finally remark that since the multiplicity of the solutions of the inversion identity is $2^{L+1}$ we have to impose the following selection rule to give account of the actual dimensionality of the transfer matrix, which is only $2^L$: 
\be r=\prod_{k=1}^L \mu_k=(-1)^{L-p}  \ee
We also introduce the following abbreviation, which will be used whenever its meaning is not ambiguous:
\be \delta_{r,1-2\textrm{Mod}(L,2)}=\delta   \ee
notice that, for even $L$
\be  \delta=\delta_{r,1} \ee
whereas for odd $L$
\be  \delta=\delta_{r,-1}  \ee

\section{Temperley Lieb Decomposition}
In this section we are going to decompose the transfer matrix $\T$ on the TL algebra. This task has carried out by analyzing the decomposition with Mathematica up to $L=9$, the general behaviour has been then recognized to hold for generic $L$.\\
We will decompose the tranfer matrix according to the following expansion:
\be \T(x) =\frac{2^{2L}}{x^L}\sum_{n=0}^{L}\D_n x^n \ee
we then introduce:
\be  {\bf H}_{2k-1}=\sqrt{2}\sum_{n=1}^{2L}[\e_n,[\e_{n+1},[\e_{n+2},\ldots[\e_{n+2k-3},\e_{n+2k-2}]\ldots]]]   \ee
\be   \mathbf{A}_2=-2L\uno  \ee
\be \mathbf{A}_4=-36L\uno   \ee
\be \mathbf{A}_6=-2400L \uno  \ee
\be \mathbf{A}_8=-352800L \uno \ee
\be \mathbf{A}_{10}=-91445760L\uno  \ee
\be  \mathbf{A}_1={\bf H}_1-2L\uno \ee
\be \mathbf{A}_3={\bf H}_{3}+6\mathbf{A}_1  \ee
\be   \mathbf{A}_5=6{\bf H}_{5}+60\mathbf{A}_3-120\mathbf{A}_1  \ee
\be  \mathbf{A}_7=90{\bf H}_{7}+210\mathbf{A}_5-5040\mathbf{A}_3+5040\mathbf{A}_1   \ee
\be  \mathbf{A}_9= 2520{\bf H}_{9}+504\mathbf{A}_7-45360\mathbf{A}_5+604800\mathbf{A}_3-362880\mathbf{A}_1    \ee
\be \mathbf{A}_{11}=113400{\bf H}_{11}+990\mathbf{A}_9-221760\mathbf{A}_7+11642400\mathbf{A}_5-99792000\mathbf{A}_3+39916800\mathbf{A}_1  \ee
and in general:
\be \mathbf{A}_{2k-1}=\frac{(2k-2)!}{2^{k-1}}{\bf H}_{2k-1}+\sum_{m=1}^{k-1}\frac{(-1)^{m+1}}{m}\binom{2k-m-2}{m-1}\frac{(2k-1)!}{(2k-2m-2)!}\mathbf{A}_{2m-1}  \ee
the first $\D_n$ then read:
\be \D_0=\uno  \ee
\be \D_1= \frac{1}{2}\mathbf{A}_1  \ee
\be  \D_2= \frac{1}{8}(\mathbf{A}_1^2+\mathbf{A}_2) \ee
\be    \D_3=\frac{1}{48}(\mathbf{A}_3+\mathbf{A}_1^3+3\mathbf{A}_2\mathbf{A}_1)   \ee
\be  \D_4= \frac{1}{384}(4\mathbf{A}_1\mathbf{A}_3+\mathbf{A}_1^4+6\mathbf{A}_2\mathbf{A}_1^2+3\mathbf{A}_2^2+\mathbf{A}_4) \ee
\be  \D_5= \frac{1}{3840}(\mathbf{A}_1^5+10\mathbf{A}_1^3\mathbf{A}_2+15\mathbf{A}_1\mathbf{A}_2^2+10\mathbf{A}_1^2\mathbf{A}_3+10\mathbf{A}_2\mathbf{A}_3+5\mathbf{A}_1\mathbf{A}_4+\mathbf{A}_5)   \ee
\be \nonumber  \ldots   \ee
\be \D_L= \frac{1}{2^{L}}\big((-1)^L\uno+\mathbf{R}\big) \ee
\be \D_{n}=\frac{1}{2^{n}n!}B_n(\mathbf{A}_1,\ldots,\mathbf{A}_n) \ee
where the $B_n(x_1,\ldots,x_n)$ are the Bell polynomials, which are generated by the following expansion:
\be \exp\Big(\sum_{n=1}^\infty \frac{x_n}{n!}t^n\Big)=1+\sum_{n=1}^\infty\frac{B_{n}(x_1,\ldots,x_n)}{n!}t^n     \ee
We call the ${\bf A}_{2n-1}$ the Lattice Local Integrals of Motion (IOM), since they play a completely analogous role as the ${\bf I}_{2n-1}$ of \cite{blz}, as we shall realize even better in the next section.\\
The ${\bf A}_{2n-1}$ form a maximal Abelian subalgebra (of hermitean operators) of the enveloping algebra of the TL algebra, in the same way as the CFT Local IOM ${\bf I}_{2n-1}$ span the maximal abelian subalgebra (of hermitean operators) of the enveloping algebra of the Virasoro algebra. As remarked in \cite{saleur} the 2 enveloping algebras should become isomorphic in a suitably defined $L\to\infty$ limit.\\

\section{Local Integrals of Motion}
In this section we are going to derive the exact expressions for the eigenvalues of the lattice Integrals of Motion and after this we will discuss how these eigenvalues are related to the eigenvalues of the conntinuum Integrals of Motion which have been already obtained by means of Themodynamic Bethe Ansatz (TBA) in \cite{nigro}, following lines advocated in \cite{tba}. In this case however no TBA will be employed, rather the eigenvalues will be recognized in an expansion that has been already carried out for Critical Dense Polymers in \cite{nigropol} by means of all order Euler-Maclaurin expansion of the factorized form of the eigenvalues. In this case however we will not employ this method, but rather direct Taylor expansion.\\
We start by reminding that
\be  t_k=\frac{(2k-\delta)}{2L}\pi   \ee
one then sees that the eigenvalues of the transfer matrix can be expanded in the following series:
\be\begin{split}  \log\Lambda^r(x)&= -L\log x+2L\log 2-\sum_{n=1}^\infty \frac{x^{2n}}{2^{2n}(2n)!}(2n-1)!2^{2n}\sum_{k=1}^L\sin^{2n}t_k+\\
&+\sum_{n=1}^\infty \frac{x^{2n-1}}{2^{2n-1}(2n-1)!}(2n-2)!2^{2n-1}\Big(2\sum_{k\in\mathcal{P}}\sin^{2n-1}t_k-\sum_{k=1}^L\sin^{2n-1}t_k\Big) \end{split} \ee 
we then recognize the eigenvalues of the operators ${\bf A}_n $:
\be A_{2n}=-(2n-1)!2^{2n}\sum_{k=1}^L\sin^{2n}t_k   \ee
and
\be  A_{2n-1}=(2n-2)!2^{2n-1}\Big(2\sum_{k\in\mathcal{P}}\sin^{2n-1}t_k-\sum_{k=1}^L\sin^{2n-1}t_k\Big)  \ee
these values of the eigenvalues are generally correct, what happens is that for $L>n-1$ the numerical diagonalization of the Temperley Lieb decomposition of ${\bf A}_{2n-1}$ given in the previous section agrees perfectly with the above expressions, below this threshold the Temperley Lieb decompositions of the previous section will degenerate to other forms, such that their eigenvalues are still given by the above expressions.\\
We choose not to study these degenerate forms since they never appear in the expansion in Bell polynomials, however the algebraic dependence of the degenerate ${\bf A}_{2n-1}$ on the lower degree IOM could be expressed by using Inverse Bell polynomials having as argument the tangles ${\bf D}_n$, of which only a finite number is different from zero.\\
We now turn our attention to the $A_{2n}$, for this purpose we are interested in the basic expressions:
\be \sum_{k=1}^L\sin^{2n}t_k  \ee  
for example one can directly check that:
\be 1!2^2\sum_{k=1}^L\sin^{2}t_k=2L ,\ L> 1  \ee
\be 3!2^4\sum_{k=1}^L\sin^{4}t_k=36L ,\ L>2 \ee
\be 5!2^6\sum_{k=1}^L\sin^{6}t_k=2400L , \ L>3  \ee
\be 7!2^8\sum_{k=1}^L\sin^{8}t_k=352800L , \ L>4  \ee
which reproduces precisely the coefficients the identity in the ${\bf A}_{2n}$, and in general:
\be A_{2n}=(2n-1)!2^{2n}\sum_{k=1}^L\sin^{2n}t_k=2n\Bigg(\frac{(2n-1)!}{n!} \Bigg)^2L , \ L>n \ee 
To deal with the eigenvalues of the ${\bf A}_{2n-1}$ we will sum up the odd powers of the sine in a degeneration independent sum of trigonometric functions, namely:
\be   \sum_{j=1}^L \sin^{2n-1}(t_j)=\frac{1}{2^{2n-2}}\sum_{k=1}^n (-1)^{k+1}\binom{2n-1}{n-k}\frac{\cos\big((\delta-1)(2k-1)\frac{\pi}{2L}\big)}{\sin\big((2k-1)\frac{\pi}{2L}\big)}         \ee
so that our analytic knowledge of the eigenvalues $A_{2n-1}$ is now completely analytic and ready for $1/L$ expansion, and recovery of the continuum IOM in the leading order behaviour.\\
We now observe that by  using the following expansion (here $B_m(z)$ are the Bernoulli polynomials):
\be  \frac{\cos((\delta-1)\frac{(2k-1)\pi}{2L})}{\sin \frac{(2k-1)\pi}{2L}}=\frac{2L}{(2k-1)\pi}+\sum_{m=1}^\infty(-1)^{m}4^m\frac{B_{2m}(1-\frac{\delta}{2})}{\Gamma(2m+1)}\Big(\frac{(2k-1)\pi}{2L}\Big)^{2m-1}   \ee
and plugging it into the formula for the sum of the odd powers of sines we get:
\be \begin{split}\sum_{j=1}^L\sin^{2n-1}(t_j)&=  \frac{L}{\pi}\frac{1}{2^{2n-3}}\sum_{k=1}^n \frac{(-1)^{k+1}}{2k-1}\binom{2n-1}{n-k}+\\
 &+\sum_{m=1}^\infty (-1)^{m}4^m\frac{B_{2m}(1-\frac{\delta}{2})}{\Gamma(2m+1)}\Big(\frac{\pi}{2L}\Big)^{2m-1} \frac{1}{2^{2n-2}}\sum_{k=1}^n(-1)^{k+1}\binom{2n-1}{n-k}(2k-1)^{2m-1} \end{split}\ee
for the contribution of the excitations we have a similar expansion
\be \begin{split}\sum_{j\in\mathcal{P}}\sin^{2n-1}(t_j)&=-\sum_{m=n}^\infty (-1)^{m}4^m\frac{m\sum_{j\in\mathcal{P}}\big(j-\frac{\delta}{2}\big)^{2m-1}}{\Gamma(2m+1)}\Big(\frac{\pi}{2L}\Big)^{2m-1} \frac{1}{2^{2n-2}}\sum_{k=1}^n(-1)^{k+1}\binom{2n-1}{n-k}(2k-1)^{2m-1} \end{split}\ee
however notice that 
\be  \sum_{k=1}^n(-1)^{k+1}\binom{2n-1}{n-k}(2k-1)^{2m-1} =0 , \quad {\rm for} \quad m<n   \ee
we now define the divergent parts of the eigenvalues $A_{2n-1}$ as:
\be \begin{split} \frac{A_{2n-1}^{div}}{(2n-2)!2^{2n-1}}&=-\frac{L}{\pi}\frac{1}{2^{2n-3}}\sum_{k=1}^n \frac{(-1)^{k+1}}{2k-1}\binom{2n-1}{n-k} \end{split}  \ee
After introducing the divergent part we can now analyze the finite part of the eigenvalues:
\be  \frac{A_{2n-1}-A_{2n-1}^{div}}{(2n-2)!2^{2n-1}}=-\sum_{m=n}^\infty (-1)^{m}4^m\frac{\alpha_m I_{2m-1}(\mathcal{P})}{\Gamma(2m+1)}\Big(\frac{\pi}{2L}\Big)^{2m-1} \frac{1}{2^{2n-2}}\sum_{k=1}^n(-1)^{k+1}\binom{2n-1}{n-k}(2k-1)^{2m-1}    \ee
being
\be \alpha_m I_{2m-1}(\mathcal{P})=2m\sum_{j\in\mathcal{P}}\big(j-\frac{\delta}{2}\big)^{2m-1}+B_{2m}\big(1-\frac{\delta}{2}\big)  \ee
and
\be  \alpha_n= -\sqrt{\pi}\frac{n(2n-1)3^n\Gamma(4n-1)}{2^{2n-2}n!\Gamma\big(3n-\frac{1}{2}\big)}                  \ee
where the $I_{2m-1}$ are the eigenvalues of the Local IOM of CFT, which have been derived in \cite{nigro}.\\

\section{Projectors, Characters and Operator Content}
Let $p$ be the length of the partition $\mathcal{P}$, one has that the projectors:
\be {\bf \Pi}^\pm=\frac{1\pm{\bf R}}{2}   \ee
project respectively on states with even or odd $p$.\\
It is interesting to mention, that due to the identities:
\be \frac{B_L(\mathbf{A}_1,\ldots,\mathbf{A}_L)}{2 L!}= \frac{\big((-1)^L\uno+\mathbf{R}\big)}{2} \ee
one obtains a TL decomposition of the projectors, which is actually very non trivial, namely, for even $L$:
\be {\bf \Pi}^+= \frac{B_L(\mathbf{A}_1,\ldots,\mathbf{A}_L)}{2 L!}   \ee
\be {\bf \Pi}^-= \uno-\frac{B_L(\mathbf{A}_1,\ldots,\mathbf{A}_L)}{2 L!}   \ee
and for odd $L$:
\be {\bf \Pi}^-= -\frac{B_L(\mathbf{A}_1,\ldots,\mathbf{A}_L)}{2 L!}   \ee
\be {\bf \Pi}^+= \uno+\frac{B_L(\mathbf{A}_1,\ldots,\mathbf{A}_L)}{2 L!}   \ee
We then consider the following characters of the projectet space of states:
\be   \chi_+^L(q,\delta)=\sum_{\mathcal{P}, \ \textrm{odd}\ p}q^{\sum_{j\in\mathcal{P}}\big(j-\frac{\delta}{2}\big)}   \ee
\be   \chi_{-}^L(q,\delta)=\sum_{\mathcal{P}, \ \textrm{even}\ p}q^{\sum_{j\in\mathcal{P}}\big(j-\frac{\delta}{2}\big)}   \ee
which, by defining the q-Binomial
\be  \binom{n}{m}_q=\prod_{i=0}^{m-1}\frac{(1-q^{n-i})}{(1-q^{i+1})}   \ee
can also be expressed in the following Fermionic form:
\be   \chi_{+}^L(q,\delta)=\sum_{m=0}^{\lfloor\frac{L}{2}\rfloor}q^{2m^2+m-m\delta}\binom{L}{2m}_q       \ee
\be   \chi_{-}^L(q,\delta)=\sum_{m=1}^{\lfloor\frac{L}{2}\rfloor+1}q^{2m^2-m-(2m-1)\frac{\delta}{2}}\binom{L}{2m-1}_q       \ee
alternatively we also have the following Bosonic forms for the finitized characters:
\be  \chi_{+}^L(q,\delta)=\frac{1}{2}\Big(\prod_{k=1}^{L}(1+q^{k-\frac{\delta}{2}})+\prod_{k=1}^{L}(1-q^{k-\frac{\delta}{2}})\Big)   \ee
\be  \chi_{-}^L(q,\delta)=\frac{1}{2}\Big(\prod_{k=1}^{L}(1+q^{k-\frac{\delta}{2}})-\prod_{k=1}^{L}(1-q^{k-\frac{\delta}{2}})\Big)   \ee
We remark that these finitized characters have been already object of extensive study, and are found, in a slightly less symmetric form for example in \cite{ret1}.\\
By use the above formulas one can see that the partition functions behave as:
\be  Z=\textrm{Tr}\Big({\bf \Pi}^{+}\T^M\Big)\sim Z_{\textrm{div}}(L,M)\chi_{+}^L(q,\delta)   \ee
and
\be  Z=\textrm{Tr}\Big({\bf \Pi}^{-}\T^M\Big)\sim Z_{\textrm{div}}(x,L,M)\chi_{-}^L(q,\delta)   \ee
where
\be   q=e^{4\pi\frac{M}{N}x}  \ee
and
\be   Z_{\textrm{div}}=\exp\Big(M\sum_{n=1}^\infty \frac{A_{2n}(L)}{2^{2n}(2n)!}x^{2n}+M\sum_{n=1}^\infty \frac{A_{2n-1}^{\textrm{div}}(L)}{2^{2n-1}(2n-1)!}x^{2n-1}\Big)   \ee
we now take a look at the expansions of the $L=\infty$ characters:
\be \chi_{+}^\infty(q,0)= 1+q^3+q^4+2q^5+2q^6+3q^7+3q^8+4q^9+5q^{10}+6q^{11}+7q^{12}+9q^{13}+11q^{14}+\ldots   \ee
\be \chi_{+}^\infty(q,1)=1+q^2+q^3+2q^4+3q^5+3q^6+5q^7+5q^8+7q^9+8q^{10}+11q^{11}+12q^{12}+16q^{13}+\ldots   \ee
\be \chi_{-}^\infty(q,0)= q+q^2+q^3+q^4+q^5+2q^6+2q^7+3q^8+4q^9+5q^{10}+6q^{11}+8q^{12}+9q^{13}+11q^{14}+\ldots   \ee
\be \chi_{-}^\infty(q,1)=q^{\frac{1}{2}}+q^{\frac{3}{2}}+q^{\frac{5}{2}}+q^{\frac{7}{2}}+2q^{\frac{9}{2}}+2q^{\frac{11}{2}}+3q^{\frac{13}{2}}+4q^{\frac{15}{2}}+5q^{\frac{17}{2}}+6q^{\frac{19}{2}}+8q^{\frac{21}{2}}+9q^{\frac{23}{2}}+\ldots    \ee
we notice that:
\be  \chi_{+}^\infty(q,1)=\chi_{0}(q)    \ee
\be  \chi_{-}^\infty(q,1)=\chi_{\frac{1}{2}}(q)    \ee
whre $\chi_0$ and $\chi_{\frac{1}{2}}$ are the respectively the characters of the irreducible $h=0$, $h=1/2$ representations of the Virasoro algebra.\\
Now, notice that the other 2 characters enumerate the number of partitions of integers into distinct parts with either an even ($\chi_+^\infty$) or an odd  ($\chi_{-}^\infty$) number of parts. We identify such characters as the characters of the irreducible $(\frac{1}{16})_\pm$ representations of conformal weight $h=\frac{1}{16}$: 
\be   \chi_{+}^\infty(q,0)=q^{-\frac{1}{16}}\chi_{\frac{1}{16}}^+(q)   \ee
\be   \chi_{-}^\infty(q,0)=q^{-\frac{1}{16}}\chi_{\frac{1}{16}}^{-}(q)   \ee
so that for even $L$ the corresponding continuum transfer matrix is decomposed into $(0)_{+}\oplus(\frac{1}{16})_{-}$ representations whereas for odd $L$ the content is $(\frac{1}{16})_{+}\oplus(\frac{1}{2})_{-}$, where the subscripts $\pm$ are used to specify the parity of the number of fermi modes which we identify with $p$.\\
The states belonging to  the $(0)_{+}$ and $(\frac{1}{2})_{-}$ representations, and labelled by a partition $\mathcal{P}$, will all be of the form:
\be  \big|\mathcal{P}\big>=\psi_{-\frac{2k_p-1}{2}}\ldots\psi_{-\frac{2k_1-1}{2}}\big|0\big>  \ee
with $p$ respectively even or odd.\\
The fermion field is expanded in this case as:
\be  \psi(z)=\sum_{m\in\mathbb{Z}+\frac{1}{2}}\frac{\psi_{m}}{z^{m+\frac{1}{2}}}  \ee
For the states belonging to the $(\frac{1}{16}_\pm)$ representation one has instead:
\be  \big|\mathcal{P}\big>=\psi_{-k_p}\ldots\psi_{-k_1}\Big|\frac{1}{16}\Big>  \ee
and for the fermion field:
\be  \psi(z)=\sum_{m\in\mathbb{Z}}\frac{\psi_{m}}{z^{m+\frac{1}{2}}}  \ee
where the fermi modes satisfy in all cases the following anticommutation relations:
\be  \{\psi_n,\psi_m\}=\delta_{n+m,0}    \ee
All the above conformal characters and fermionic decompositions are well known and appear in many works, we point out here for example \cite{paths}\cite{ginsparg} as particularly understandable.
\section{Conclusions}
In this paper we have obtained a complete decomposition of the transfer matrix for the Ising model with periodic boundaries on the periodic TL algebra. The program of building a lattice version of the Local IOM of \cite{blz} has been completely carried out, and the precise relation of the lattice IOM to the continuum IOM is given through a $1/L$ expansion whose coefficients are proportional to the continuum IOM. Unsurprisingly we find that the operator content of the model is classified by the irreducible representations of the $\mathcal{M}_{3,4}$ CFT.\\

\end{document}